\documentclass[runningheads,fleqn]{svmult}
\usepackage{makeidx}   
\usepackage{graphicx}  
\usepackage{subeqnar}  
\usepackage{multicol}  
\usepackage{taphys}    
\makeindex             
\begin{document}
\title*{Cohesion and Stability of Metal Nanowires: A Quantum Chaos 
Approach\footnote{B.\ Kramer (Ed.): Adv. in Solid State Phys.\ {\bf 41}, 
497-511 (2001)}}
\toctitle{Cohesion and Stability of Metal Nanowires: A Quantum Chaos Approach}
%
%
\titlerunning{Cohesion and Stability of Metal Nanowires} 
%
\author{C.\ A.\ Stafford\inst{1}
\and F.\ Kassubek\inst{2}
\and Hermann Grabert\inst{2}}
\authorrunning{C.\ A.\ Stafford et al.}
%
%
\institute{Department of Physics, University of Arizona\\
1118 E.\ 4th Street, Tucson, AZ 85721, USA
\and Fakult\"at f\"ur Physik, Albert-Ludwigs-Universit\"at\\
Hermann-Herder-Stra\ss e  3, D-79104 Freiburg, Germany}

\maketitle              

\begin{abstract}
\index{abstract} 
A remarkably quantitative understanding 
of the electrical and mechanical properties of 
metal wires with a thickness on the scale of a nanometer
has been obtained within the free-electron model 
using semiclassical techniques.  
Convergent trace formulas for the density of states and
cohesive force of a narrow constriction in an electron gas, whose 
classical motion is either chaotic or integrable, are derived.  
Mode quantization in a metallic point contact or nanowire leads to 
universal oscillations in its cohesive force, whose amplitude 
depends only on a dimensionless quantum parameter describing the crossover
from 
chaotic to integrable motion, and is of order 1nN, in agreement with 
experiments on gold nanowires.  A linear stability analysis shows that 
the classical instability of a long wire under surface tension can be
completely
suppressed by quantum effects, leading to stable cylindrical 
configurations whose electrical conductance is a magic number 1, 3, 5, 6,... 
times $2e^2/h$, in accord with recent results on alkali metal 
nanowires.
\end{abstract}

\section{Introduction}
\label{sec:intro}

In 1971, Gutzwiller's trace formula \cite{Trace}
expressing the quantum density
of states of a classically chaotic system as a Feynman sum over classical
periodic orbits gave birth to the field of quantum chaos.
In the subsequent decades, the trace formula was generalized, and 
applied to a wide variety of physical systems \cite{Gutzwiller,SemiclPhys}.
Of particular interest here are trace formulas for systems with 
continuous symmetries \cite{Balian,CreaghPRA90} and broken symmetries
\cite{UllmoPRE96,CreaghAnnPhys96}.  One of the most 
important successes of this semiclassical
approach has been the description of 
shell effects in finite fermion systems
\cite{SemiclPhys,Meier}.  In this article, we discuss the application
\cite{Hoeppler99,StaffordPRL99,Kassubek00,Kassubek01} 
of trace formulas 
to describe quantum-size effects in a particular class
of open quantum systems: metallic nanocontacts and nanowires.

\begin{figure}[t]
\hspace*{-4mm}
\resizebox{6cm}{!}{
\includegraphics*[11cm,54mm][20cm,18cm]{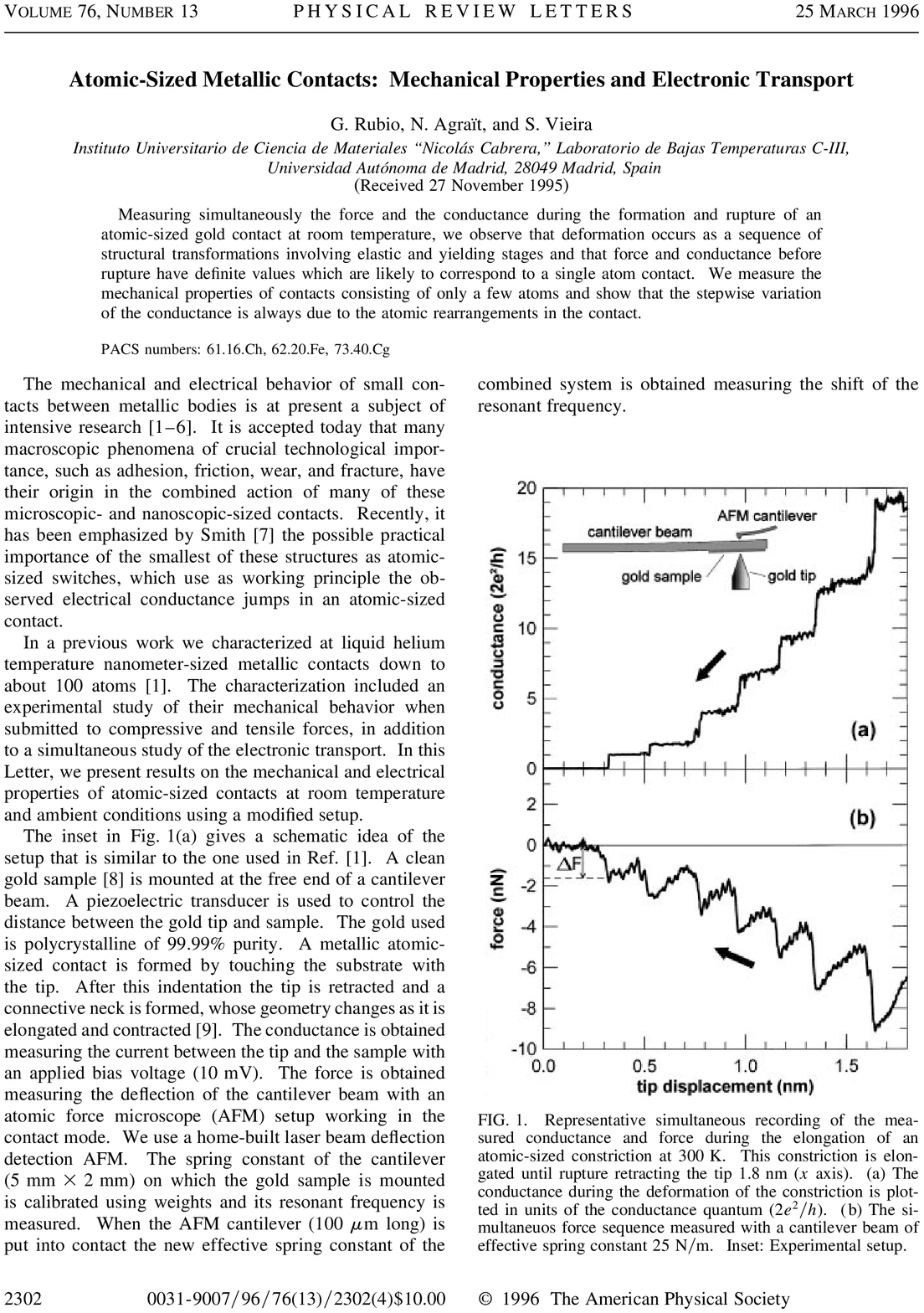}}
\hspace*{-4mm}
\resizebox{6.0cm}{!}{
\includegraphics*{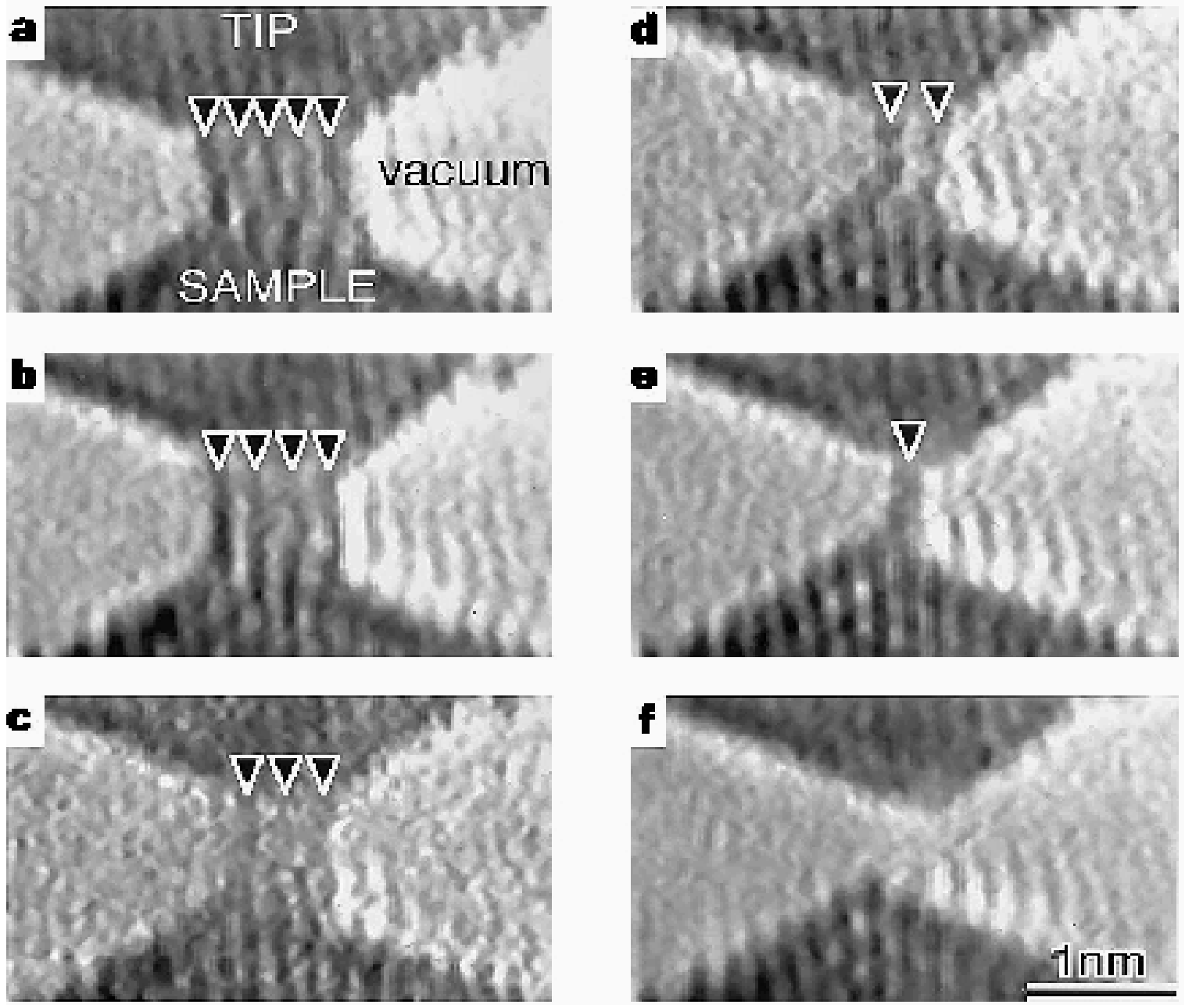}}
\caption{
Left: Simultaneous measurements
of ({\bf a}) the conductance and ({\bf b})
the cohesive force of a gold nanowire during elongation
at room temperature, from Ref.\  \cite{Agrait}.  
Right: Transmission electron micrographs
of an atomic-scale gold contact breaking, from
Ref.\  \cite{Auchain}.  The measured electrical
conductance of the contact is 
({\bf d}) $G\simeq 2G_0$, ({\bf e}) $G\simeq G_0$
}
\label{fig1}
\end{figure}

In the past eight years, experimental
research on 
atomically-thin metal wires has burgeoned
\cite{Nanowires,Agrait,Durig,bridge,liquid,Scheer,Auchain,Auchain2,Yanson,Kondo}.
In a seminal experiment \cite{Agrait}
carried out in 1995, Rubio, Agra\"{\i}t and Vieira
simultaneously measured the electrical conductance and cohesive force of
an atomic-scale gold contact as it formed and ruptured
(see Fig.\ \ref{fig1}).
They observed steps of order $G_0=2e^2/h$ in the conductance, which were
synchronized with a sawtooth structure with an amplitude 
of order 1nN in the force.  Similar results were obtained independently by
Stalder and D\"urig \cite{Durig}.  For comparison, electron
micrographs by Ohnishi {\em et al.} \cite{Auchain}
illustrating the atomic-scale structure 
of a gold nanocontact breaking are also shown in Fig.\ \ref{fig1}.

Conductance steps of size $G_0$ were first observed in quantum point contacts
(QPCs)
fabricated in semiconductor heterostructures \cite{beenakker},
and are a rather universal phenomenon in metal nanowires \cite{Nanowires}, 
even being found in contacts formed in liquid metals \cite{liquid}.  
The precision of conductance
quantization in metal nanocontacts is poorer than that in semiconductor
QPCs due to their inherently rough structure on the scale of the 
Fermi wavelength $\lambda_F$, which causes backscattering \cite{BuerkiPRB99},
and due to the imperfect hybridization of the atomic orbitals in the
contact, especially for multivalent atoms \cite{Scheer}.
As we shall see in the following, the sawtooth structure in the cohesive
force can be considered 
a mechanical analogue of conductance quantization
\cite{StaffordPRL97}.

\begin{figure}[t]
\sidecaption
\hspace*{-12mm} 
\resizebox{9cm}{!}{
\includegraphics*[11cm,4cm][200mm,85mm]{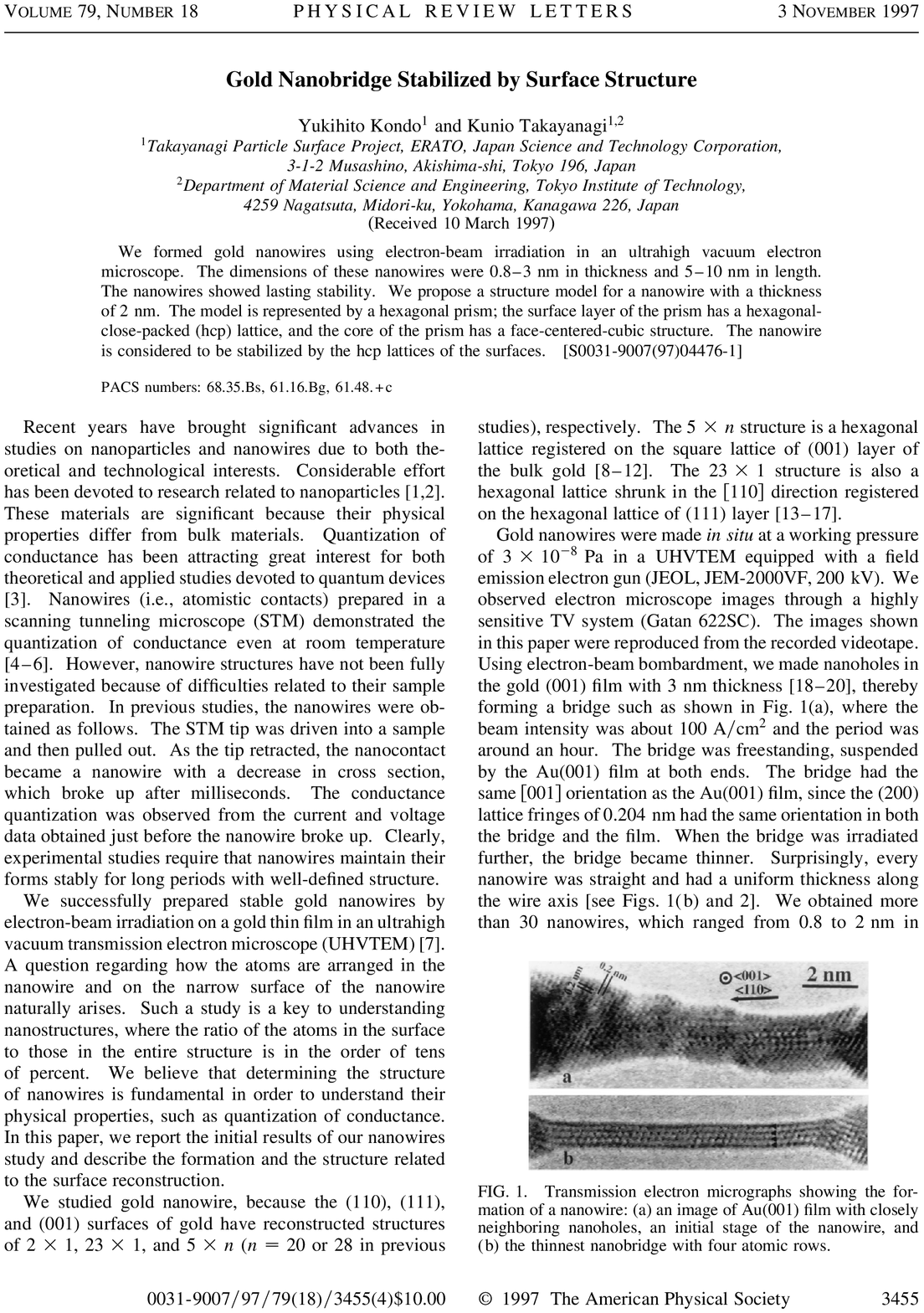}}
\hspace*{-9mm}
\caption{Transmission electron micrographs showing the formation of a
gold nanowire, from Ref.\ \cite{bridge}: ({\bf a}) an image of Au(001)
film with closely spaced nanoholes, the initial stage of the nanowire;
({\bf b}) a nanowire four atoms in diameter, resulting from further
electron-beam irradiation}
\label{fig3}
\end{figure}

A remarkable feature of metal nanowires is the fact that they are stable at
all.  Fig.\ \ref{fig3} shows electron micrographs by Kondo and Takayanagi
\cite{bridge} illustrating the formation
of a gold nanowire.  Under electron beam irradiation, the wire becomes
ever thinner, until it is but four atoms in diameter.  Almost all of the 
atoms are at the surface, with small coordination numbers. The surface energy
of such a structure is enormous, yet it is observed to form spontaneously, and
to persist almost indefinitely.  Even wires one atom thick, such as that
shown in Fig.\ \ref{fig1}(e), are found to be stable for days at a time
\cite{Auchain,Auchain2}.  Naively, such structures might be expected to 
break apart due to surface tension \cite{Plateau,Chandrasekhar,Powers},
but we shall show that quantum-size effects can stabilize arbitrarily
long nanowires \cite{Kassubek01}.

\section{Free electron model}
\label{sec:free}

\begin{figure}[t]
\sidecaption
\hspace*{-12mm}
\resizebox{9cm}{!}{
\includegraphics*{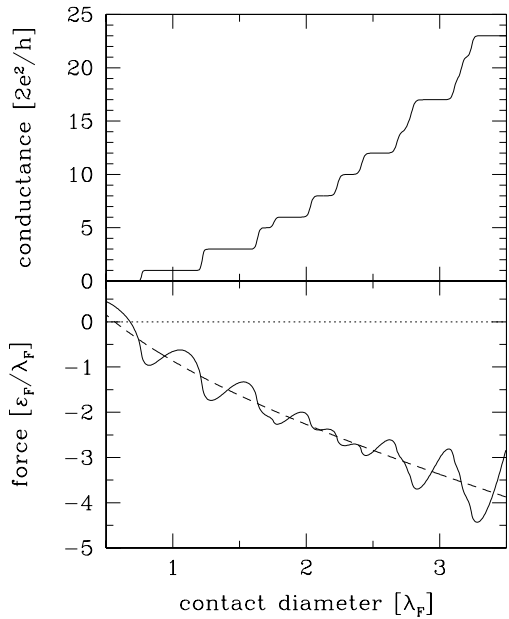}}
\hspace*{-9mm}
\caption{Electrical conductance and cohesive force of a nanowire, modeled as
a narrow neck in a free-electron gas, calculated from Eqs.\ 
(\ref{eq:landauer})--(\ref{eq:dos_dmb}) at zero temperature.
The $S$-matrix was calculated using the adiabatic
and WKB approximations, following Ref.\ \cite{StaffordPRL97}. 
For comparison, the contribution to the force from the surface tension
and curvature energy is shown as a dashed line.  Note that
$\varepsilon_F/\lambda_F=1.7\mbox{nN}$ in gold}
\label{fig_fandg}
\end{figure}

We investigate the simplest possible model 
\cite{StaffordPRL97,Kassubek99} for a metal
nanowire: a free (conduction) electron gas confined within the wire by
Dirichlet boundary conditions.  A nanowire is an open quantum system, and
so is treated most naturally in terms of the electronic scattering matrix
$S$.  The Landauer formula \cite{Landauer57,FisherPRB81}
expressing the electrical conductance in terms of the submatrix
$S_{12}$ describing transmission through the wire is
\begin{equation}
G=\frac{2e^2}{h} \int dE \frac{-\partial f(E)}{\partial E} \mbox{Tr}\left\{
S_{12}^\dagger(E) S_{12}(E)\right\},
\label{eq:landauer}
\end{equation}
where $f(E)$ is the Fermi-Dirac distribution function.
The conductance of a metal nanocontact was calculated exactly in this model
by Torres {\em et al.} \cite{Torres}.
The appropriate thermodynamic potential to describe the energetics of 
such an open system is the grand canonical potential $\Omega$, whose 
derivative with respect to elongation gives the cohesive force $F$:
\begin{equation}
\Omega = -\frac{1}{\beta} \int dE\, g(E) \ln \left(1+e^{-\beta(E-\mu)}\right),
\;\;\;\;\;\;\;\;\;
F=-\frac{\partial \Omega}{\partial L}.
\label{eq:grandpot}
\end{equation}
Here $\beta$ is the inverse temperature, $\mu$ is the 
chemical potential of electrons injected into the nanowire from the
macroscopic electrodes, and $g(E)$ is the 
electronic density of states (DOS) of the nanowire.  The DOS of an open
system may be expressed in terms of the scattering matrix as
\cite{Dashen}
\begin{equation}
g(E) = \frac{1}{2\pi i} \mbox{Tr} \left\{S^{\dagger}(E)\frac{\partial S}{
\partial E} - \mbox{H.c.}
\right\}.
\label{eq:dos_dmb}
\end{equation}
This formula is also known as the Wigner delay.  
Note that in Eqs.\ (\ref{eq:landauer}) and (\ref{eq:dos_dmb}),
a factor of 2 for spin degeneracy has been included. 
Thus, once the electronic
scattering problem for the nanowire is solved, both the conductance and
force can be readily calculated \cite{StaffordPRL97,Kassubek99,BuerkiPRB99},
as shown in
Fig.\ \ref{fig_fandg}.
One sees that there is an almost quantitative agreement with the experimental
results shown in Fig. \ref{fig1}: for example, the force necessary to
break the last conducting channel is approximately
$\varepsilon_F/\lambda_F$ (=1.7nN in 
gold), where $\varepsilon_F$ is the Fermi energy.

\section{Weyl expansion}
\label{sec:weyl}

In order to separate out the mesoscopic sawtooth structure in the force,
associated with the opening of individual conductance channels, from the
overall (macroscopic) 
trend of the contact to become stronger as its diameter increases, it is
useful to perform a systematic semiclassical expansion 
\cite{Gutzwiller,SemiclPhys} of the DOS,
$g(E) = \bar{g}(E) + \delta g(E)$,
where $\bar{g}$ is a smooth average term, referred to as the
Weyl contribution, and $\delta g(E)$ is an oscillatory term, whose average
is zero.  For the free electron model with Dirichlet boundary conditions,
the Weyl term is \cite{SemiclPhys}
\begin{equation}
\bar{g}(E) = 
E^{-1}\left(\frac{k_E^3 V}{2\pi^2} 
- \frac{k_E^2 A}{8\pi} 
+ \frac{k_E K}{6\pi^2} \right),
\label{eq:weyl2}
\end{equation}
where $k_E=\sqrt{2mE}/\hbar$,
$V$ is the volume of the wire, $A$ its 
surface area, and $K$ the integrated mean curvature of its surface.
Inserting Eq.\ (\ref{eq:weyl2}) into Eq.\
(\ref{eq:grandpot}), one finds the following semiclassical expansion
 at zero temperature:
\begin{equation}
\frac{\Omega}{\varepsilon_F}  =  
-\frac{2k_F^3 V}{15\pi^2} + \frac{k_F^2 A}{16\pi} -
\frac{2k_F K}{9\pi^2} + \frac{\delta \Omega}{\varepsilon_F}.
\label{eq:weyl_ener}
\end{equation}
One can show \cite{StaffordPRL99}
that interaction effects are higher order in $\hbar$.
In the same spirit, a semiclassical expansion for
the conductance  
$G=(2e^2/h)G_S+\delta G$ may be derived,
where the corrected Sharvin conductance is \cite{Torres}
\begin{equation}
G_S =\left(\frac{k_F D^*}{4}\right)^2
\left(1 -\frac{4}{k_F D^*}\right).
\end{equation}
Here $D^*$ is the narrowest diameter of the nanowire.

When the wire is elongated, the atoms rearrange themselves, but the volume
per atom remains essentially constant \cite{StaffordPRL99,Kassubek99}.  Thus,
when differentiating Eq.\ (\ref{eq:weyl_ener}) to calculate $F$, the
first term on the r.h.s.\ gives zero:
\begin{equation}
F = -\frac{\partial \Omega}{\partial L}
=  -\sigma \frac{\partial A}{\partial L} + 
\gamma \frac{\partial K}{\partial L}
 + \delta F.
\end{equation}
The cohesive force is given by 
surface tension plus a small curvature correction (the sum of which
is indicated by a dashed curve in Fig.\ \ref{fig_fandg}),
combined with an oscillatory quantum term.

\section{Trace formulas}
\label{sec:trace}

The oscillatory contribution $\delta g(E)$ to the DOS
may be approximated as a Feynman sum
over classical periodic orbits \`a la Gutzwiller \cite{Gutzwiller,SemiclPhys}.
Since we are interested in modeling nanowires which may possess axial and/or
translational symmetries, however, we can not in general utilize Gutzwiller's 
original trace formula \cite{Trace}, which describes systems whose periodic
orbits are isolated, but must instead employ a generalization due
to Creagh and Littlejohn, describing a system with an $f$-dimensional Abelian
symmetry \cite{CreaghPRA90}:
\begin{equation}
\delta g(E)  = 
\frac{2}{\pi \hbar} \frac{1}{(2 \pi \hbar)^{f/2}} 
\sum_\Gamma \frac{ T_\Gamma V_\Gamma J_\Gamma^{-1/2}}{
|\det \tilde{M}_{\Gamma}-1|^{1/2}}
\cos\left(\frac{S_\Gamma}{\hbar}- \frac{\sigma_\Gamma
\pi}{2} -  \frac{f\pi}{4}
\right),
\label{eq:trace_sym}
\end{equation}
where the sum runs over $f$-dimensional 
families $\Gamma$ of degenerate periodic orbits,
$T_\Gamma$ is the period of an orbit in $\Gamma$,
$V_\Gamma$ is the $f$-dimensional volume spanned by $\Gamma$, 
$S_\Gamma$ is the action of the orbit, and
$\sigma_\Gamma$ is a phase shift determined by the singular points along the 
classical trajectory.
The quantity $\tilde{M}$ is the so-called monodromy
matrix, characterizing the stability of the orbit with respect to
perturbations. It describes as a Poincar{\'e} map the linearized
motion of small perturbations from the periodic orbit in a surface of section
perpendicular to the orbit in phase space:
an initial variation of momentum and position in the surface of
section $(\delta r, \delta p)$ is related to the  mismatch
$(\delta r', \delta p')$ after one period by
\begin{equation}
\label{eq:def_of_monodromi_matrix}
\left( \begin{array}{c} \delta r' \\ \delta p'\end{array} \right) =
\tilde{M} \left( \begin{array}{c} \delta r \\ \delta p\end{array} \right).
\end{equation}
Finally, the factor
$J_\Gamma=|\det(\partial r'/\partial p)|$.

We shall also need to consider the breaking of continuous symmetries,
which is elegantly described in terms of
semiclassical perturbation theory \cite{UllmoPRE96,CreaghAnnPhys96},
wherein the cosine in the trace formula is replaced by
\begin{equation}
\cos(S_\Gamma/\hbar + \theta_\Gamma) \rightarrow \mbox{Re}
\left\{ e^{i(S_\Gamma/\hbar 
+\theta_\Gamma)}
\left\langle
e^{i\Delta S_\Gamma/\hbar} 
\right\rangle_\Gamma
\right\},
\label{eq:symbreak1}
\end{equation}
where
\begin{equation}
\langle e^{i\Delta S_\Gamma/\hbar} \rangle_\Gamma = V_\Gamma^{-1} \!
\int
d\mu(g)e^{i\Delta S_\Gamma(g)/\hbar}
\label{eq:symbreak2}
\end{equation}
is an average over the measure of the broken symmetry group.

\subsection{A 2D example}
\label{subsec:2D}

\begin{figure}[t]
\resizebox{9cm}{!}{
\includegraphics*{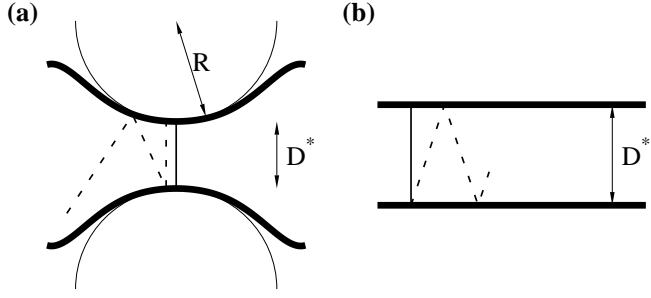}}
\caption{Point contact ({\bf a}) and straight wire ({\bf b}) 
as limiting cases of
a nanowire. The point contact is characterized by the diameter
$D^*$ and radius of curvature $R$ of the neck. For
the straight wire, $R\rightarrow \infty$. A periodic orbit is sketched
with a solid line, other orbits (dotted lines) are not periodic. 
The classical motion in the point contact ({\bf a}) is in general chaotic, 
while
the straight wire ({\bf b}) possesses integrable classical motion
}
\label{fig_2dcontact}
\end{figure}

Before treating the three-dimensional problem of interest, it is 
instructive to consider a two-dimensional analogue, which is much simpler, but
already contains the essential elements of the problem.
To be specific, we consider a QPC whose width varies as
\begin{equation}
D(z) = D^* + z^2/R, \;\;\; z \in [-L/2,L/2]
\end{equation}
along the wire (see Fig.\ \ref{fig_2dcontact}).
For a finite radius of curvature $R$, there is only a single unstable
periodic orbit (plus harmonics), which moves up and down at the narrowest
point of the neck.
The monodromy matrix $\tilde{M}_{\rm ppo}$
of the primitive periodic orbit is given by
\begin{equation}
\label{eq:semi:Mhalbe}
\tilde{M}_{\rm ppo}^{1/2}=\left.\left( \begin{array}[c]{cc}
\frac{ \partial r'}{\partial r}
& \frac{ \partial r'}{\partial p}
\vspace*{2mm} \\
\frac{ \partial p'}{\partial r}
& \frac{ \partial p'}{\partial p}
                       \end{array}
\right)\right|_{\frac{1}{2}{\rm ppo}}
 =
\left(
\begin{array}[c]{cc}
1+D^*/R & \hspace*{3mm}
D^*(1+D^*/2R)/p 
\vspace*{2mm}\\
2 p /R 
& 1+D^*/R
\end{array}
\right),
\end{equation}
with eigenvalues
\begin{equation}
\label{eq:semi:xi2D}
e^{\pm\chi}= 1+D^*/R \pm \sqrt{(1+D^*/R)^2-1},
\end{equation}
$2\chi$ being the Lyapunov exponent of the primitive periodic orbit.
There is no continuous symmetry present ($f=0$), so the original
Gutzwiller trace formula \cite{Trace} may be used to find \cite{StaffordPRL99}
\begin{equation}
\delta g_0(E)=
\frac{ 2 mD^* }{\pi \hbar^2 k_E}
\sum_{n=1}^\infty \frac{\cos(2n k_E D^*)}{|\sinh(n \chi)|}.
\label{eq:trace0}
\end{equation}

In the limit $R \rightarrow \infty$, the Lyapunov exponent $\chi\rightarrow 0$,
and Eq.\ (\ref{eq:trace0}) diverges.  In this limit, the wire acquires
translational symmetry ($f=1$), and Eq.\ (\ref{eq:trace_sym}) may be used to
find
\begin{equation}
\frac{\delta g_1(E)}{L} = \frac{2m D^*}{\pi\hbar^2} \sum_{n=1}^\infty
\frac{\cos(2n k_E D^* - \pi/4)}{\sqrt{\pi n k_E D^*}}.
\label{eq:trace1}
\end{equation}
The classical motion is integrable in this limit.

For large but finite radii of curvature, one can employ
semiclassical perturbation theory in $R^{-1}$:
\begin{equation}
\langle e^{i\Delta S_n/\hbar} \rangle_z = 
\frac{ 1}{L \sqrt{D^*}}\int_{-L/2}^{L/2} dz\, D(z)^{1/2}
e^{-i2nk(E)[D^*-D(z)]}.
\end{equation} 
Ignoring the $1/R$-dependence of the prefactor, one finds
\begin{equation}
\langle e^{i\Delta S_n/\hbar} \rangle_z = \frac{
C(\sqrt{n k(E) L^2/R\pi})+iS(\sqrt{n k(E) L^2 /R\pi})}{
\sqrt{n k(E)L^2/R\pi}},
\end{equation}
where $C$ and $S$ are Fresnel integrals.
This leads to a DOS
\begin{equation}
\delta g_{\rm pert}(E)  = 
\frac{2 m D^*}{\pi \hbar^2 k_E} 
\sum_{n=1}^\infty \frac{
{\cal C}\left(2 n k_E D^*- \frac{\pi}{4},
\sqrt{\frac{n k_E L^2}{\pi R}}\right) 
}{n\sqrt{D^*/R}},
\label{eq:trace_pert}
\end{equation} 
where we have defined the function
\begin{equation}
{\cal C}(x,y)\equiv \cos(x) {\rm C}(y) - \sin(x){\rm S}(y).
\end{equation}

The Gutzwiller formula (\ref{eq:trace0}) may be expanded in a Taylor
series around $R=0$, while the perturbation formula (\ref{eq:trace_pert})
gives a Laurent series around $R=\infty$.  Combining the two, 
an interpolation formula valid for arbitrary $R$ can be constructed 
\cite{StaffordPRL99}:
\begin{equation}
\delta g_{\rm int}(E)  = 
\frac{ \sqrt{8} 
m D^*}{\pi \hbar^2 k_E} \sum_{n=1}^\infty \frac{
{\cal C}\left(2 n k_E D^*- \frac{\pi}{4},
\sqrt{\frac{n k_E L^2}{\pi R}}\right) 
}{|\sinh(n \chi)|}.
\label{eq:trace_int}
\end{equation}
The crossover from integrable to chaotic behavior in 
Eq.\ (\ref{eq:trace_int}) is controlled by the
dimensionless parameter
\begin{equation}
\alpha(E) = L/\sqrt{\lambda_E R},
\end{equation}
where $\lambda_E=2\pi/k_E$ is the de Broglie wavelength of an electron of
energy $E$.  We refer to $\alpha$ as the quantum chaos parameter: 
for $\alpha \ll 1$ the DOS is indistinguishable
from that of an integrable system, while
for $\alpha \gg 1$, the DOS is that of a chaotic system.

\begin{figure}[t]
\sidecaption
\hspace*{-8mm}
\resizebox{9cm}{!}{
\includegraphics{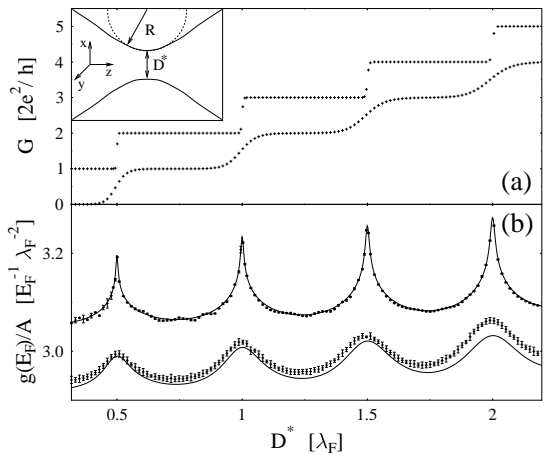}}
\hspace*{-1cm}
\caption{({\bf a}) Conductance $G$ and ({\bf b}) DOS $g(E_F)$ for 2D
nanocontacts with $\alpha \approx 5$ versus the contact diameter
$D^*$. $g$ is normalized to the area $A$ of the region. Solid curves:
semiclassical results based on the interpolation formula; crosses with
error bars: numerical results obtained by a recursive Green's function
method \cite{BuerkiPRB99}. 
Lower curves in ({\bf a}) and ({\bf b}): $R \approx \lambda_F$; upper
curves (offset vertically): $ R\approx 170 \lambda_F$
}
\label{fig:comparison}
\end{figure}

In Fig.\ \ref{fig:comparison}, 
the DOS calculated from Eq.\ (\ref{eq:trace_int})
plus the 2D Weyl term is compared to the result of a numerical
solution of the Schr\"odinger equation.
Remarkably, the semiclassical
result is seen to be quantitatively accurate even in the extreme quantum
limit $D^* \sim \lambda_F$, $R \sim \lambda_F$.

\subsection{3D nanowire with axial symmetry}
\label{subsec:3D}

\begin{figure}[t]
\sidecaption
\resizebox{6cm}{!}{
\includegraphics*{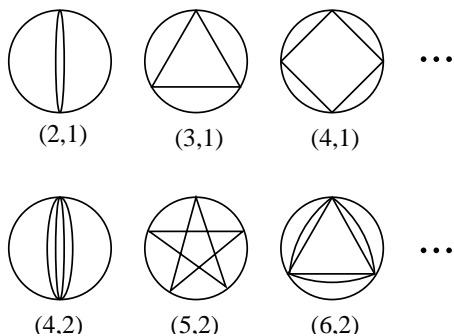}}
\caption{
Periodic orbits of an electron in the narrowest cross-section of the 
neck, labeled $(v,w)$, where $v$ is the number of vertices and
$w$ the winding number.  The length of an orbit is 
$L_{vw}=v D^* \sin \phi_{vw}$, where 
$\phi_{vw}=\pi w/v$ is the angle of incidence at a vertex
}
\label{fig_orbits}
\end{figure}

\begin{figure}[t]
\sidecaption
\hspace*{-4mm}
\resizebox{8cm}{!}{
\includegraphics{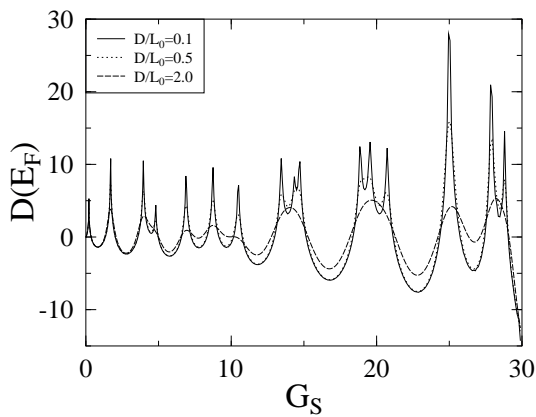}}
\hspace*{-.5cm}
\caption{DOS at the Fermi energy for  axially symmetric 3D
nanowires of parabolic shape versus the Sharvin conductance $G_S$. The
different curves represent contacts with various ratios of $D/L_0$
(indicated in the inset), where  $D$ is the asymptotic
diameter of the constriction and $L_0$ its initial
length
}
\label{fig:semi:dos3D_interpol}
\end{figure}

For an axially-symmetric three-dimensional nanocontact, the periodic orbits
(see Fig.\ \ref{fig_orbits}) occur in one-dimensional families which
fit into the narrowest cross-section of the contact.
This problem was first investigated by Balian and Bloch
\cite{Balian}, who derived the axially-symmetric analogue of Eq.\ 
(\ref{eq:trace0}).  We can follow the
procedure outlined in Sec.\ \ref{subsec:2D} to derive an interpolation
formula describing the crossover from a long nanowire ($f=2$) to a short
nanocontact ($f=1$) \cite{StaffordPRL99}:
\begin{equation}
\delta g(E)=
\frac{ m}{\hbar^2 
} \sum_{w=1}^\infty \sum_{v=2 w}^\infty
\frac{f_{vw} L_{vw}^{3/2}
{\cal C}\left(k_E L_{vw}-3v\pi/2, \alpha(E)\sqrt{v\sin\phi_{vw}}
\right)}{v^2 |\sinh(v \chi_{vw}/2)| \sqrt{\pi k_E}},
\label{eq:trace_int3D}
\end{equation}
where
$f_{vw} = 1 + \theta(v-2w)$ counts the discrete symmetry of the orbit under
time-reversal,  the Lyapunov exponent $\chi_{vw}$ is given by
\begin{equation}
e^{\chi_{vw}} =  
1+\frac{L_{vw} \sin \phi_{vw}
}{vR}+\sqrt{\left(1+\frac{ L_{vw} \sin \phi_{vw}}{vR}\right)^2-1},
\end{equation}
and the remaining terms are defined in
the caption of Fig.\ \ref{fig_orbits}.  Eq.\ (\ref{eq:trace_int3D})
is plotted in Fig.\ \ref{fig:semi:dos3D_interpol}.  Note the rounding of
the peaks in the DOS in short contacts.

\section{Universal force oscillations}

\begin{figure}[t]
\vspace*{-3.5cm}
\resizebox{10cm}{!}{
\includegraphics*{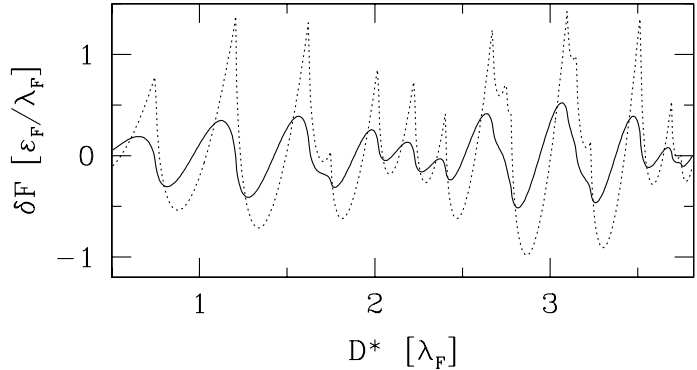}}
\vspace*{-1.3cm}
\caption{Force oscillations $\delta F$ versus the minimum contact
diameter $D^*$: dashed curve: $\lim_{\alpha\rightarrow 0}\{\delta F\}$;
solid curve: $\lim_{\alpha \rightarrow \infty} \{\alpha \delta F\}$.
The result for $\alpha \gg 1$ is consistent with the WKB calculation
shown in Fig.\ \ref{fig_fandg}, while the result for $\alpha \rightarrow
0$ (integrable limit) agrees with the result \cite{Hoeppler99} for
a straight wire}
\end{figure}

The characteristic amplitude of
the sawtooth structure in the cohesive force of a gold nanocontact was found
to be of order 1nN, 
independent of the contact area or shape \cite{Agrait,Durig}.
To what extent is the
amplitude of the force oscillations {\em universal}?
To calculate the force from Eq.\ (\ref{eq:grandpot}), we need to make some
assumptions regarding how the shape of the contact scales under elongation.
First, we assume that the deformation occurs primarily in the thinnest
section, which implies $D^{*2}L\approx \mbox{const}$.  Second, we assume
that $R\propto L^2$, which implies $\alpha =L/\sqrt{\lambda_F R}\approx
\mbox{const}$.  Inserting Eq.\ (\ref{eq:trace_int3D}) into Eq.\ 
(\ref{eq:grandpot}), and taking the derivative, we find \cite{StaffordPRL99}:
\begin{equation}
\delta F 
\begin{array}{c}\mbox{ }\\ \simeq \\ {\scriptstyle \alpha \gg 1} \end{array}
-\frac{\varepsilon_F}{L} \sum_{w=1}^{\infty}
\sum_{v=2w}^{\infty} \sqrt{\frac{L_{vw}}{\lambda_F}} \frac{f_{vw}
\sin(k_F L_{vw}-3v\pi/2 + \pi/4)}{v^2\sinh(v\chi_{vw}/2)}, 
\end{equation}
\begin{equation}
\delta F \begin{array}{c} \mbox{ }\\ \simeq  \\ {\scriptstyle \alpha \ll 1}
 \end{array} 
-\frac{2\varepsilon_F}{\lambda_F} \sum_{w=1}^{\infty} \sum_{v=2w}^{\infty}
\frac{f_{vw}}{v^2} \sin(k_F L_{vw}-3v\pi/2). 
\end{equation}
\begin{equation}
\mbox{rms}\,
\delta F
= \frac{\varepsilon_F}{\lambda_F}\times\left\{\begin{array}[c]{l r} 
0.58621, & \alpha \ll 1,\\
0.36208 \, \alpha^{-1}, & 
\alpha \gg 1.
\end{array}\right.
\end{equation}
From Figs.\ \ref{fig1} and \ref{fig3}, one sees that $\alpha < 1$ for a
realistic geometry, implying that indeed $\mbox{rms}\,\delta F \sim 
\varepsilon_F/\lambda_F$.

\section{Quantum suppression of the Rayleigh instability}

\begin{figure}[h]
\sidecaption
\resizebox{7cm}{!}{
\includegraphics*{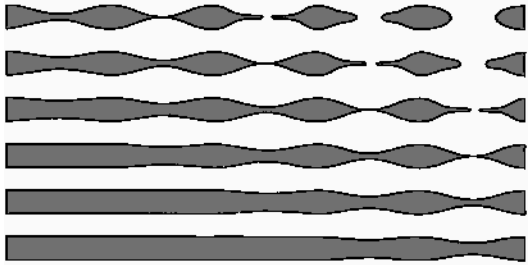}}
\caption{Artist's conception of a propagating Rayleigh instability, 
from Ref.\ \cite{Powers}}
\label{fig:powers}
\end{figure}

A cylindrical body longer than its circumference is unstable to breakup
under surface tension \cite{Plateau,Chandrasekhar} 
(see Fig.\ \ref{fig:powers}).  How then to explain the durability of long
gold nanowires [c.f.\ Fig.\ \ref{fig3}(b)], the thinnest of
which have been shown \cite{Kondo} 
to be almost perfectly cylindrical in shape?  
Let us calculate the quantum corrections 
\cite{Kassubek01} to the classical stability analysis.
Classically, only axially-symmetric deformations 
lead to instabilities.  Any such
deformation of a cylinder may be written as a Fourier series
\begin{equation}
R(z)=R_0 +\int_{-\infty}^\infty dq\, b(q) e^{iq z},
\end{equation}
where $R(z)$ is the radius of the cylinder at $z$ and $R_0$ the unperturbed 
radius.  The constant volume constraint leads to the condition
\begin{equation}
b(0)=-\frac{ 1}{ R_0} \int_0^\infty dq\, |b(q)|^2.
\label{eq:constraint}
\end{equation}
In the Weyl approximation, the energy cost of the deformation is
\begin{eqnarray}
\label{eq:stab:omegaWeyl}
\Delta \bar{\Omega}/\varepsilon_F&=&
        \left(-\frac{8}{15} k_F^3 R_0+\frac{\pi}{4} k_F^2 \right) b(0)
\nonumber \\&&
        + \int_0^\infty dq\,\left[-\frac{8k_F^3}{15}
+ \left(\frac{\pi k_F^2 R_0}{4} -\frac{ 8 k_F}{9}\right)  q^2
\right] |b(q)|^2.
\end{eqnarray}
For the unperturbed cylinder, Eq.\ (\ref{eq:trace_sym}) yields
\begin{equation}
\delta g(E)=\frac{mL}{\pi\hbar^2}\sum_{w=1}^\infty \sum_{v=2w}^\infty
\frac{f_{vw}L_{vw}}{v^2} \cos(k_E L_{vw}-3v\pi/2).
\end{equation}
The effect of the deformation may be treated with
semiclassical perturbation theory:
\begin{equation}
\langle e^{i\Delta S_{vw}(z)/\hbar} \rangle_z = 
\frac{1}{LR_0}
\int_0^L dz \,R(z)\, e^{i\Delta S_{vw}(z)/\hbar},
\end{equation}
where
\begin{equation}
\frac{\Delta S_{vw}(z)}{\hbar}
 = 2v \sin \phi_{vw} k_E 
\int_{-\infty}^\infty dq\, b(q) e^{iq z}.
\end{equation}
Expanding $\delta g$ up to second order in $b(q)$ gives
\begin{eqnarray}
\Delta\{\delta g(E)\} & = &
\frac{4m}{\hbar^2} \sum_{w=1}^\infty \sum_{v=2w}^\infty
\frac{f_{vw} \sin\phi_{vw}}{v}
 \left[b(0) 
(\cos\theta_{vw}
 -k_E L_{vw}\sin\theta_{vw})\rule{0mm}{5mm}\right.
\nonumber \\
&&\mbox{} 
- \left.\frac{k_E L_{vw}}{R_0}\int_0^\infty \! dq\, |b(q)|^2
\left(\sin\theta_{vw}
+ \frac{k_E L_{vw}}{2}\cos\theta_{vw}\right)
\right],
\label{eq:long}
\end{eqnarray}
where $\theta_{vw}(E)=k_E L_{vw}-3v\pi/2$.  

\begin{figure}[t]
\sidecaption
\hspace*{-5mm}
\resizebox{8.5cm}{!}{
\includegraphics*{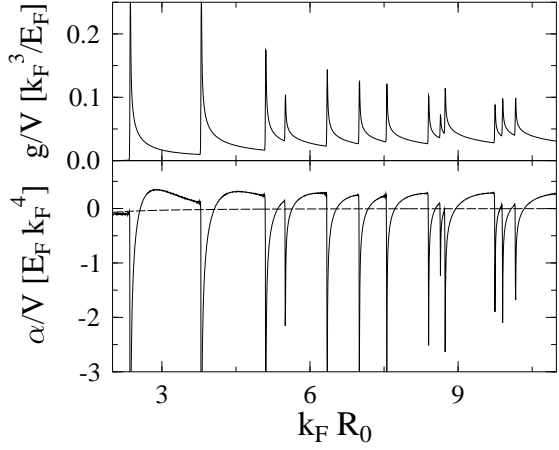}} 
\hspace*{-4mm}
\caption{Density of states $g(E_F)$ of a cylindrical wire
(upper diagram) and
stability coefficient $\alpha$ (lower diagram) versus the
radius $R_0$ of the unperturbed wire.
The wavevector of the perturbation is $q R_0 =1$.
Dashed curve: Weyl contribution to $\alpha$
}
\label{fig:alpha}
\end{figure}

Combining Eqs.\ 
(\ref{eq:stab:omegaWeyl}) and (\ref{eq:long}), and using the 
constraint (\ref{eq:constraint}), one finds that 
the change of the DOS is of second order in $b$,
and contributions with different $q$ decouple.  
The energy integral (\ref{eq:grandpot}) yields
\begin{equation}
\Omega[b]=\Omega[0]+\int_0^\infty\!\! dq\, \alpha(q) |b(q)|^2
+ {\cal{O}}(b^3),
\end{equation} 
where the stability coefficient $\alpha(q)$ depends implicitly on
$R_0$ and temperature.
If $\alpha(q)$ is
negative for any value of $q$, then $\Omega$ decreases under the deformation
and the wire is unstable.

\begin{figure}[t]
\sidecaption
\hspace*{-3mm}
\resizebox{8.3cm}{!}{
\includegraphics*{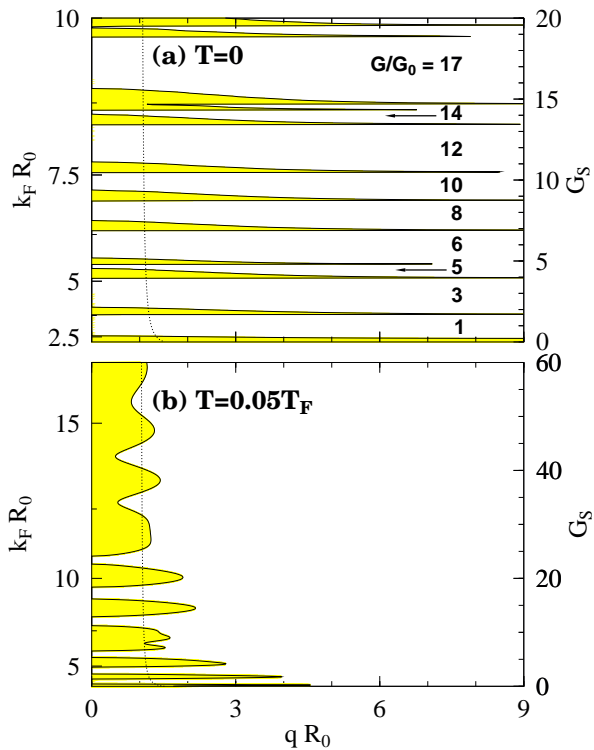}}
\caption{Stability diagram for cylindrical nanowires at two different
temperatures.
White areas are stable, grey unstable to small perturbations.
The quantized electrical conductance values $G$ of the stable
configurations are indicated by bold numerals in ({\bf a}), with $G_0=2e^2/h$.
Right vertical axis: corrected Sharvin conductance $G_S$.
Dotted curve: stability criterion in the Weyl approximation
}
\label{xx:stab:fig2}
\end{figure}

Fig.\ \ref{fig:alpha} shows the stability coefficient and DOS at the 
classical stability threshold $qR_0=1$ as a function of $R_0$.  The quantum
correction 
destabilizes the wire where the DOS is sharply peaked; but what is more
surprising, it {\em stabilizes} 
the wire in the intervening intervals.  With these results, we can construct
a stability diagram for the wire.  For a given temperature, the stability 
problem is now determined by two dimensionless parameters: $qR_0$ and
$k_FR_0$.  In Fig.\ \ref{xx:stab:fig2}, regions of instability, where
$\alpha(q)<0$, are shaded grey, while stable regions are shown in white.
Note that many of the white regions of stability persist all the way down to
$q=0$, indicating that an infinitely long wire is a true metastable state
if its radius lies in one of the windows of stability.
The quantized conductance values of the stable cylindrical
configurations are indicated by bold numerals in Fig.\ \ref{xx:stab:fig2}(a).
Our stability analysis is consistent with recent experimental results
for alkali metal nanowires \cite{Yanson}.

\section*{Acknowledgments}

CAS is indebted to Dionys Baeriswyl and J\'er\^ome B\"urki for their 
contributions to the early phase of this work, and to Raymond Goldstein
for his insights on the quantum Rayleigh problem.
CAS was supported by NSF Grant DMR0072703.
FK and HG were supported by Grant
SFB 276 of the Deutsche Forschungsgemeinschaft. 
This research was supported by an award from Research Corporation.


%

\end{document}